\documentclass[12pt, letterpaper, preprint, comicneue]{aastex63}
\usepackage[T1]{fontenc}
\usepackage{fontawesome}
\usepackage{color}
\usepackage{amsmath}
\usepackage{natbib}
\usepackage{ctable}
\usepackage{bm}
\usepackage[normalem]{ulem} 
\usepackage{xspace}
\usepackage{paralist}

\let\oldbibliography\thebibliography 
\renewcommand{\thebibliography}[1]{%
  \oldbibliography{#1}%
  \setlength{\itemsep}{0pt}%
  \setlength{\parsep}{0pt}%
  \setlength{\parskip}{0pt}%
  \setlength{\bibsep}{0ex}
  \raggedright
}
\newcommand{\given}{\,|\,}

\newcommand{\bfi}[1]{\textbf{\textit{#1}}}

\newcommand{\eg}{\emph{e.g.}}

\let\oldAA\AA
\renewcommand{\AA}{\text{\normalfont\oldAA}}

\newcommand{\btheta}{\boldsymbol{\theta}}
\newcommand{\bphi}{\boldsymbol{\phi}}

\newcommand{\foreign}[1]{\textsl{#1}}
\newcommand{\etal}{\foreign{et~al.}}

\newcommand{\bitem}{\begin{itemize}}
\newcommand{\eitem}{\end{itemize}}
\newcommand{\beq}{\begin{equation}}
\newcommand{\eeq}{\end{equation}}

\newcommand{\github}{\href{https://github.com/changhoonhahn/SEDflow/}{\faGithub}}

\newcommand{\sedflow}{{\sc SEDflow}}

\definecolor{orange}{rgb}{1,0.5,0}

\begin{document} \sloppy\sloppypar\frenchspacing 

\title{Accelerated Bayesian SED Modeling using Amortized Neural Posterior Estimation}

\newcounter{affilcounter}
\author[0000-0003-1197-0902]{ChangHoon Hahn}
\altaffiliation{changhoon.hahn@princeton.edu.com}
\affil{Department of Astrophysical Sciences, Princeton University, Princeton NJ 08544, USA} 

\author[0000-0002-8873-5065]{Peter Melchior}
\affil{Department of Astrophysical Sciences, Princeton University, Princeton NJ 08544, USA} 
\affil{Center for Statistics and Machine Learning, Princeton University, 
Princeton, NJ 08544, USA}

\begin{abstract}
    State-of-the-art spectral energy distribution (SED) analyses use a
    Bayesian framework to infer the physical properties of galaxies from
    observed photometry or spectra.
    They require sampling from a high-dimensional space of SED model parameters
    and take $>10-100$ CPU hours per galaxy, which renders them practically
    infeasible for analyzing the {\em billions} of galaxies that will be
    observed by upcoming galaxy surveys (\eg~DESI, PFS, Rubin, Webb, and Roman).
    In this work, we present an alternative scalable approach to 
    rigorous Bayesian inference using Amortized Neural Posterior
    Estimation (ANPE). 
    ANPE is a simulation-based inference method that employs neural networks
    to estimate the posterior probability distribution over the full
    range of observations.
    Once trained, it requires no additional model evaluations to estimate the
    posterior.  
    We present, and publicly release, \sedflow, an ANPE method to produce
    posteriors of the recent \cite{hahn2022} SED model from optical
    photometry.
    \sedflow~takes \emph{${\sim}1$ second per galaxy} to obtain the posterior
    distributions of 12 model parameters, all of which are in excellent
    agreement with traditional Markov Chain Monte Carlo sampling results.
    We also apply \sedflow~to 33,884 galaxies in the NASA-Sloan Atlas and
    publicly release their posteriors.
    \github
\end{abstract}
\keywords{galaxies: evolution -- galaxies: statistics}

\section{Introduction} \label{sec:intro} 
Physical properties of galaxies are the building
blocks of our understanding of galaxies and their evolution. 
We can determine properties such as stellar mass ($M_*$), star formation rate (SFR), metallicity
($Z$), and age ($t_{\rm age}$) of a galaxy by analyzing its
spectral energy distribution (SED).
Theoretical modeling of galaxy SEDs is currently based on stellar population
synthesis (SPS) and describes the SED as a composite stellar population
constructed
from isochrones, stellar spectra, an initial mass function (IMF), a star
formation and chemical evolution history, and dust
attenuation~\citep[\emph{e.g.}][see \citealt{walcher2011, conroy2013} for a
comprehensive review]{bruzual2003, maraston2005, conroy2009}.
Some models also include dust and nebular emissions as well as emissions from
active galactic nuclei~\citep[\emph{e.g.}][]{johnson2021}.
In state-of-the-art SED modeling, theoretical SPS models are compared to
observed SEDs using Bayesian inference, which accurately quantifies parameter
uncertainties and degeneracies among them~\citep{acquaviva2011,
chevallard2016, leja2017, carnall2018, johnson2021, hahn2022}. 
The Bayesian approach also enables marginalization over nuisance parameters,
which are necessary to model the effects of observational systematics
(\emph{e.g.} flux calibration).

However, current Bayesian SED modeling methods, which use Markov Chain Monte
Carlo (MCMC) sampling techniques, take $10-100$ CPU hours per
galaxy~\citep[\emph{e.g.}][]{carnall2019a, tacchella2021}. 
While this is merely very expensive with current data sets of hundreds of
thousands of galaxy SEDs, observed by the Sloan Digital Sky
Survey~\citep[SDSS;][]{york2000}, DEEP2~\citep{davis2003},
COSMOS~\citep{scoville2007}, and GAMA~\citep{baldry2018}, it is prohibitive for
the next generation of surveys.
Over the next decade, surveys with the 
Dark Energy Spectroscopic Instrument~\citep[DESI;][]{desicollaboration2016},
the Prime Focus Spectrograph~\citep[PFS;][]{takada2014}, 
the Vera C. Rubin Observatory~\citep{ivezic2019}, 
the James Webb Space Telescope~\citep{gardner2006},
and the Roman Space Telescope~\citep{spergel2015}, will observe \emph{billions}
of galaxy SEDs.
The task of SED modeling alone for these surveys would amount to tens or
hundreds of billions of CPU hours, exceeding \emph{e.g.} the entire compute
allocation of the Legacy Survey of Space and Time (LSST) data release
production\footnote{$\approx$2 billion core hours
(\url{https://dmtn-135.lsst.io/})} by at least two orders of magnitude.
Recently, \cite{alsing2020} adopted neural emulators to accelerate SED model
evaluations by three to four orders of magnitude --- posterior inference takes
minutes per galaxy.
While this renders current data sets within reach, the next generation data
sets will still require tens or hundreds of millions of CPU hours whenever any
aspect of the SED model is altered.
Furthermore, this still practically precludes rapid analyses of upcoming
transient surveys, especially LSST, which will report $\sim$10,000 alerts per
minute\footnote{\url{https://dmtn-102.lsst.io/}}.

But Bayesian inference does not require MCMC sampling.  
Simulation-based inference (SBI) is a rapidly developing class of inference
methods that offers alternatives for many applications~\citep[see][and
references therein]{cranmer2020}.
Many SBI methods leverage the latest developments in statistics and Machine
Learning for more efficient posterior estimation~\citep{papamakarios2017,
alsing2019a, hahn2019c, dax2021, huppenkothen2021, zhang2021}. 
Of particular interest for SED modeling is a technique called Amortized
Neural Posterior Estimation (ANPE). 
Instead of using MCMC to sample the posterior for every single galaxy
separately, ANPE uses neural density estimators (NDE) to build a model of the
posterior for \emph{all} observable galaxies.
Once the NDE is trained, generating the posterior requires only the observed
SED and no additional model evaluations.

In this work, we present \sedflow, a method that applies ANPE to Bayesian
galaxy SED modeling using the recent \cite{hahn2022} SED model. 
We demonstrate that we can derive accurate posteriors with \sedflow~and make
Bayesian SED modeling fully scalable for the billions of galaxies that will be
observed by upcoming surveys.
As further demonstration, we apply \sedflow~to the optical photometry of
${\sim}33,000$ galaxies in the NASA-Sloan Atlas~(NSA). 
We begin in Section~\ref{sec:sbi} by describing SBI using ANPE.
We then present how we design and train \sedflow~in Section~\ref{sec:sedflow}
and describe the NSA observations in Section~\ref{sec:obs}. 
We validate the accuracy of the posteriors from \sedflow~in
Section~\ref{sec:results} and discuss the implications of our results in
Section~\ref{sec:discuss}. 

\section{Simulation-Based Inference} \label{sec:sbi}
The goal of Bayesian SED modeling, and probabilistic inference more
broadly, is to infer the posterior probability distributions
$p(\btheta\given\bfi{x})$ of galaxy properties, $\btheta$, given observations, 
$\bfi{x}$.
For a specific $\btheta$ and $\bfi{x}$, we typically evaluate the posterior
using Bayes' rule, 
$p(\btheta\given\bfi{x}) \propto p(\btheta)~p(\bfi{x}\given\btheta)$, where 
$p(\btheta)$ denotes the prior distribution and $p(\bfi{x}\given\btheta)$ the
likelihood, which is typically assumed to have a Gaussian functional form: 
\beq
\label{eq:likelihood}
    \ln p(\bfi{x}\given\btheta) = -\frac{1}{2}\left(\bfi{x} - m(\btheta)\right)^T {\bf C}^{-1}
    \left(\bfi{x} - m(\btheta)\right).
\eeq
$m(\btheta)$ is the theoretical model, in our case a galaxy SED model from SPS.
${\bf C}$ is the covariance matrix of the observations. 
In practice, off-diagonal terms are often ignored and measured uncertainties
are used as estimates of the diagonal terms. 

Simulation-based inference (SBI; also known as ``likelihood-free'' inference)
offers an alternative that requires no assumptions about the form of the
likelihood. 
Instead, SBI uses a generative model, \emph{i.e.} a simulation $F$, to generate
mock data $\bfi{x}'$ given parameters $\btheta'$: $F(\btheta') = \bfi{x}'$. 
It uses a large number of simulated pairs $(\btheta', \bfi{x}')$ to directly estimate
either the posterior  $p(\btheta\given \bfi{x})$, the likelihood
$p(\bfi{x}\given \btheta)$, or the joint distribution of the parameters and data $p(\btheta, \bfi{x})$. 
SBI has already been successfully applied to a number of Bayesian parameter
inference problems in astronomy~\citep[\emph{e.g.}][]{cameron2012, weyant2013,
hahn2017b, kacprzak2018, alsing2018, wong2020, huppenkothen2021, zhang2021}
and in physics~\citep[\emph{e.g.}][]{brehmer2019, cranmer2020}.


\subsection{Amortized Neural Posterior Estimation} \label{sec:flow}
SBI provides another a critical advantage over MCMC inference methods --- it
enables \emph{amortized inference}. 
For SED modeling using MCMC, each galaxy requires >$10^5$ model evaluations to
accurately estimate $p(\btheta \given \bfi{x})$~(\citealt{hahn2022}, 
Kwon~\etal~in prep.).
Moreover, model evaluations for calculating the posterior of one galaxy cannot
be used for another. 
This makes MCMC approaches for SED modeling of upcoming surveys computationally
infeasible.

With density estimation SBI, we require a large number (${\sim}10^6$) of model
evaluations only initially to train a neural density estimator (NDE), a neural
network with parameters $\bphi$ that is trained to estimate the density
$p_\phi(\btheta \given \bfi{x}')$.
If the training covers the entire or the practically relevant portions of the
$\btheta$ and $\bfi{x}$ spaces, we can evaluate
$p_\phi(\btheta\given\bfi{x}_i)$ for each galaxy $i$ with minimal computational
cost. 
The inference is therefore amortized and no additional model evaluations are
needed to generate the posterior for each galaxy.
This technique is called  Amortized Neural Posterior Estimation (ANPE) 
and has recently been applied to a broad range of astronomical applications
from analyzing gravitational waves~\citep[\emph{e.g.}][]{wong2020,dax2021} to
binary microlensing lensing~\citep{zhang2021}.
For SED modeling, the choice in favor of using ANPE is easy: the entire upfront
cost for ANPE model evaluations would only yield posteriors of tens of galaxies
with MCMC.

ANPE makes two important assumptions.
First, the simulator $F$ is capable of generating mock data $\bfi{x}'$ that is
practically indistinguishable from the observations.
In terms of the expected signal, $m$ in Eq.~\ref{eq:likelihood}, this is the
same requirement as any probabilistic modeling approach. 
But unlike likelihood-based evaluations, such as conventional MCMC, data
generated for SBI need to include all relevant noise terms as well. 
We address both aspects in Sections \ref{sec:training} and \ref{sec:forward-model}.
Second, ANPE assumes that the NDE is well trained: 
$p_\phi(\btheta \given \bfi{x}')$ is a good approximation of 
$p(\btheta \given \bfi{x}')$, and therefore of $p(\btheta \given \bfi{x})$. 
We assess this in Section~\ref{sec:results}.

ANPE commonly employs so-called ``normalizing flows''~\citep{tabak2010,
tabak2013} as density estimators.
Normalizing flow models use an invertible bijective transformation, $f$, to map
a complex target distribution to a simple base distribution, $\pi(\bfi{z})$, that is
fast to evaluate.
For ANPE, the target distribution is $p(\btheta \given \bfi{x})$ and the
$\pi(\bfi{z})$ is typically a simple multivariate Gaussian, or mixture of Gaussians.
The transformation $f: \bfi{z} \rightarrow \btheta$ must be invertible and have a
tractable Jacobian. 
This is so that we can evaluate the target distribution from $\pi(\bfi{z})$ by
a change of variable:  
\begin{equation} \label{eq:normflow}
    p(\btheta \given \bfi{x}) = \pi(\bfi{z}) \Bigl|{\rm det} \left(\frac{\partial
    f^{-1}}{\partial \btheta} \right)\Bigr|.
\end{equation} 
Since the base distribution is easy to evaluate, we can also easily evaluate
the target distribution.  
A neural network is trained to obtain $f$ and the collection of its parameters
form $\bphi$.
The network typically consists of a series of simple transforms (\emph{e.g.}
shift and scale transforms) that are each invertible and whose Jacobians are
easily calculated. 
By stringing together many such transforms, $f$ provides an extremely flexible
mapping from the base distribution.

Many different normalizing flow models are now available in the
literature~\citep[\emph{e.g.}][]{germain2015, durkan2019}.
In this work, we use Masked Autoregressive
Flow~\citep[MAF;][]{papamakarios2017}. 
The autoregressive design~\citep{uria2016} of MAF is particularly well-suited
for modeling conditional probability distributions such as the posterior. 
Autoregressive models exploit chain rule to expand a joint probability of a set
of random variables as products of one-dimensional conditional
probabilities: $p(\bfi{x}) = \prod_i p(x_i\given x_{1:i-1})$. 
They then use neural networks to describe each conditional probability,
$p(x_i\given x_{1:i-1})$. 
In this context, we can add a conditional variable $y$ on both sides of the
equation, $p(\bfi{x}\given \bfi{y}) = \prod_i p(x_i\given x_{1:i-1}, \bfi{y})$, so that the
autoregressive model describes a conditional probability $p(\bfi{x}\given \bfi{y})$. 
One drawback of autoregressive models is their sensitivity to the ordering of
the variables. 
Masked Autoencoder for Distribution Estimation~\citep[MADE;][]{germain2015}
models address this limitation using binary masks to impose the autoregressive
dependence and by permutating the order of the conditioning variables.
A MAF model is built by stacking multiple MADE models.  
Hence, it has the autoregressive structure of MADE but with more flexibility to
describe complex probability distributions.  
In practice, we use the MAF implementation in the $\mathtt{sbi}$ Python
package\footnote{\url{https://github.com/mackelab/sbi/}}~\citep{greenberg2019,
tejero-cantero2020}.

\pagebreak
\section{SEDflow} \label{sec:sedflow}
In this section, we present \sedflow, which applies ANPE to galaxy SED modeling
for a scalable and accelerated approach.
For our SED model, we use the state-of-the-art PROVABGS model from
\cite{hahn2022}. 
Although many SED models have been recently used in the
literature~(\emph{e.g.} {\sc Bagpipes}, \citealt{carnall2018}; 
{\sc Prospector}, \citealt{leja2017, johnson2021}), we choose PROVABGS because
it will be used to analyze >10 million galaxy spectrophotometry measured by the
DESI Bright Galaxy Survey~(Hahn~\etal~in prep.).
Below, we describe the PROVABGS model, the construction of the
\sedflow~training data using PROVABGS, and the training procedure for \sedflow.

\subsection{SED Modeling: PROVABGS} \label{sec:provabgs}
We use the state-of-the-art SPS model of the
PROVABGS~\citep{hahn2022}. 
The SED of a galaxy is modeled as a composite of stellar populations defined by
stellar evolution theory (in the form of isochrones, stellar spectral
libraries, and an initial mass function) and its star
formation and chemical enrichment histories (SFH and ZH), attenuated by
dust~\citep[see][for a review]{walcher2011, conroy2013}. 
The PROVABGS model, in particular, utilizes a non-parametric SFH with a
starburst, a non-parametric ZH that varies with time, and a flexible dust
attenuation prescription.

The SFH has two components: one based on non-negative matrix factorization
(NMF) bases and the other, a starburst component.
The SFH contribution from the NMF component is a linear combination of four NMF
SFH basis functions, derived from performing NMF~\citep{lee1999, cichocki2009,
fevotte2011} on SFHs of galaxies in the Illustris cosmological hydrodynamical
simulation~\citep{vogelsberger2014, genel2014, nelson2015}.
The NMF SFH prescription provides a compact and flexible representation of the
SFH.
The second starburst component consists of a single stellar population (SSP)
and adds stochasticity to the SFH. 

The ZH is similar defined using two NMF bases dervied from Illustris. 
This ZH prescription enables us to flexibly model a wide range of ZHs and,
unlike most SED models, it does not assume constant metallicity over time,
which can significantly bias inferred galaxy properties~\citep{thorne2021}. 
The stellar evolution theory is based on Flexible Stellar Population
Synthesis~\citep[FSPS;][]{conroy2009, conroy2010c} with the MIST
isochrones~\citep{paxton2011, paxton2013, paxton2015, choi2016, dotter2016},  
the \cite{chabrier2003} initial mass function (IMF), and a combination of the
MILES~\citep{sanchez-blazquez2006} and BaSeL~\citep{lejeune1997, lejeune1998,
westera2002} libraries.
The SFH and ZH are binned into 43 logarithmically-space time and SSPs are
evalulated at each time bin using FSPS. 
The SSPs are summed up to get the unattenuated rest-frame galaxy SED. 

Lastly, PROVABGS attenuates the light from the composite stellar population
using the two component \cite{charlot2000} dust attenuation model with
diffuse-dust (ISM) and birth cloud (BC) components. 
All SSPs are attenuated by the diffuse dust using the \cite{kriek2013}
attenuation curve.
Then, the BC component provides extra dust attenuation on SSPs younger than 100
Myr with young stars that are embedded in modecular clouds and HII regions. 
In total the PROVABGS SED model has 12 parameters: stellar mass ($M_*$),
six SFH parameters ($\beta_1, \beta_2, \beta_3, \beta_4, t_{\rm burst}, f_{\rm
burst}$), two ZH parameters ($\gamma_1, \gamma2$), and three dust attenuation
parameters ($\tau_{\rm BC}, \tau_{\rm ISM}, n_{\rm dust}$). 
Each PROVABGS model evaluation takes ${\sim}340$ ms.

\begin{figure}
\begin{center}
\includegraphics[width=0.85\textwidth]{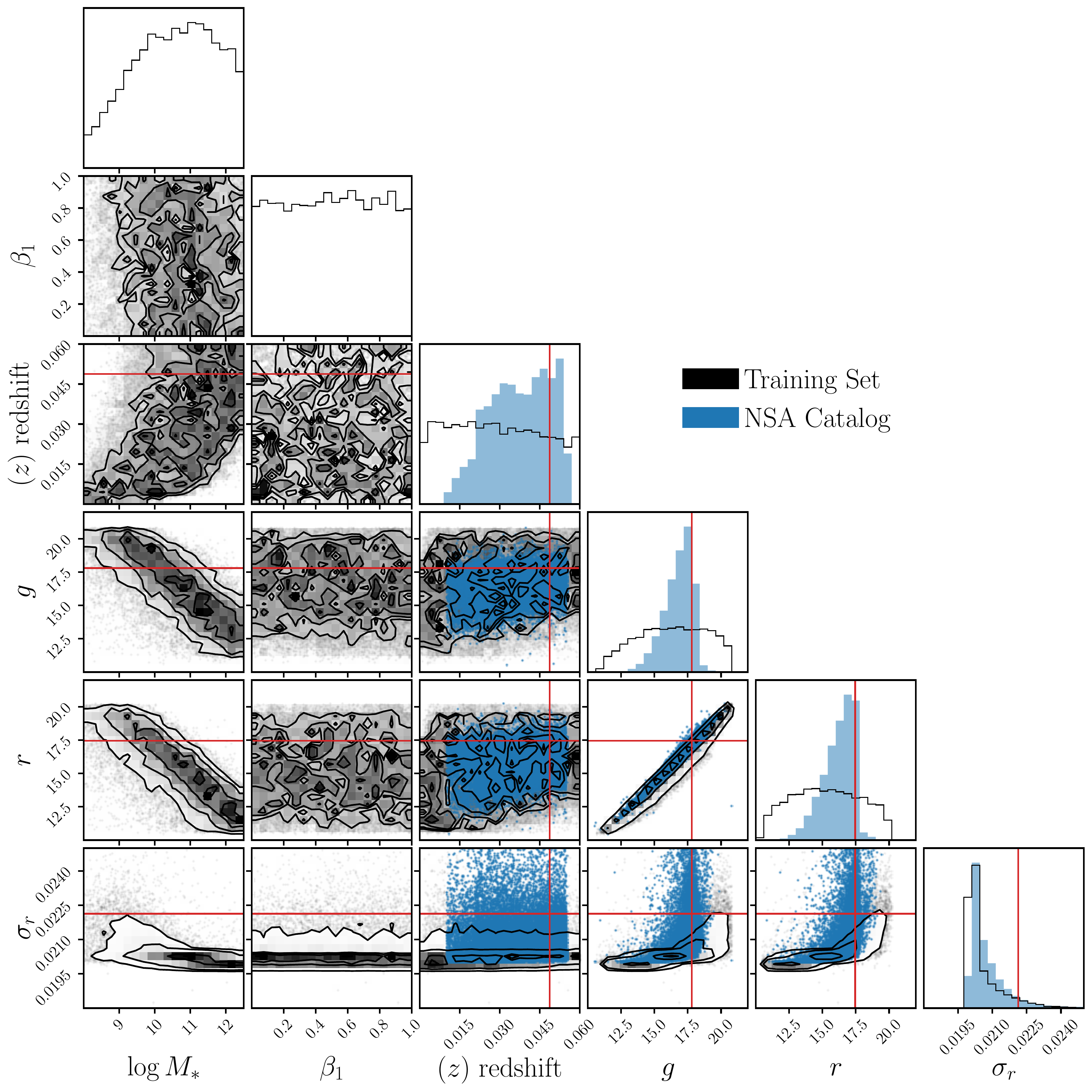}
    \caption{\label{fig:data}
    The distribution of SED model parameters, redshift, photometric
    magnitudes, and uncertainties of the data used to train \sedflow.
    We plot a subset of the parameters ($\log M_*$, $\beta_1$) and photometric
    bands for clarity.
    The training data was constructed by sampling SED model parameters from the
    prior and forward modeling optical photometry using the PROVABGS and noise
    models (Section~\ref{sec:sedflow}). 
    For comparison, we present the distribution of redshift, magnitudes, and
    uncertainties for galaxies in the NSA catalog (blue). 
    \emph{The training set encompasses the observations; thus, {\sc SEDflow}
    can be used to infer the posterior for the NSA galaxies}.
    }
\end{center}
\end{figure}
\subsection{Training Data} \label{sec:training}
In this section, we describe how we construct the training data for
\sedflow~using the PROVABGS SED model.
First, we sample $N_{\rm train}$ model parameters from a prior: $\theta'\sim p(\theta)$. 
We use the same priors as \cite{hahn2022}: uniform priors over $M_*,
t_{\rm burst}, f_{\rm burst}, \gamma_1, \gamma_2, \tau_{\rm BC}, \tau_{\rm ISM},
n_{\rm dust}$ with broad conservative ranges and Dirichlet prior over $\beta_1,
\beta_2, \beta_3, \beta_4$, chosen to normalize the NMF SFH.
For each $\theta'$, we also uniformly sample a redshift within the range of the
NSA: $z' \sim \mathcal{U}(0., 0.2)$. 
Next, we forward model mock observables. 
We calculate the rest-frame galaxy SED from PROVABGS and redshift it: 
$F(\lambda;\theta', z)$. 
Afterwards, we convolve $F$ with optical broadband filters, $R_X$, to generate
noiseless photometric fluxes:
\begin{equation}
    f_X(\theta', z') = \int F(\lambda;\theta', z') \, R_X(\lambda) \, {\rm d}\lambda
\end{equation}
The next step of the forward model is to apply noise. 
We assign photometric uncertainties, $\sigma'_X$, by sampling an estimate of
the observed $p(\sigma_X | f_X)$ of NSA galaxies. 
Then, we apply Gaussian noise
\begin{equation} \label{eq:noise} 
    \hat{f}_X(\theta', z', \sigma'_x) = f_X(\theta', z') + n_X  \quad {\rm where}~n_X \sim \mathcal{N}(0, \sigma'_X)
\end{equation}
to derive the forward modeled photometric flux.

For our estimate of $p(\sigma_X | f_X)$, we use an empirical estimate based on
NSA photometry and measured uncertainties. 
For each of the five optical bands, we separately estimate  
\begin{equation}
    \hat{p}(\sigma_X | f_X) = \mathcal{N} \big( \mu_{\sigma_X}(f_X),
    \sigma_{\sigma_X}(f_X) \big)
\end{equation}
as a Gaussian in magnitude-space. 
$\mu_{\sigma_X}$ and $\sigma_{\sigma_X}$ are the median and standard deviation
of $\sigma_X$ as a function of $f_X$ that we estimate by evaluating them in
$f_X$ bins and interpolating over the bins. 
Any $\theta'$ that is assigned a negative $\sigma'_X$ is removed from our
training data. 
We also remove any training data with $f_X(\theta')$ outside the range of NSA
photometry. 

As SBI requires an accurate noise model, one might be concerned about the
simplicity of our model.
If the noise model is incorrect, estimates of the parameters $\theta$ would be
biased by an amount that is impossible to predict.
We therefore add the noise variances $\sigma_X$ to the conditioning variables
of the posterior model, \emph{i.e.} we train and evaluate 
$p_\phi(\btheta\given\bfi{x},\{\sigma_X\})$.
This means we merely have to ensure that $\sigma'_X$ spans the observed
$\sigma_X$ values in order to have a posterior that is robust to our choice of
noise model.
We discuss this further in Section~\ref{sec:discuss}.

In total, we constuct $N_{\rm train} = 1,131,561$ sets of SED parameters,
redshift, photometric uncertainties, and mock NSA photometry. 
In Figure~\ref{fig:data}, we present the distribution of the training data
$\{(\theta', z, \sigma'_X, \hat{f}_X) \}$ (black).
We include select SED model parameters ($\log M_*$, $\beta_1$), redshift,
photometry in the $g$ and $r$ bands, and photometric uncertainty in the $r$
band.
We also include the $(z, \sigma_X, f_X)$ distribution of NSA galaxies (blue),
for comparison.
The photometry and uncertainties are in magnitude-space. 
The distribution of the training data spans the distribution of NSA galaxies.

\subsection{Training \sedflow} \label{sec:anpe_train}
For \sedflow, we use a MAF normalizing flow model (Section~\ref{sec:flow}) with 
15 MADE blocks, each with 2 hidden layers and 500 hidden units.
In total, the model has 7,890,330 parameters, $\bphi$. 
We determine this architecture through experimentation. 
Our goal is to determine $\bphi$ of the MAF model, 
$p_\phi(\btheta\given\bfi{x})$, so that it accurately estimates the
posterior probability distribution $p(\btheta\given\bfi{x})$. 
$\btheta$ represent the SED parameters and $\bfi{x} = (f_X, \sigma_X, z)$.
We do this by minimizing the KL divergence between 
$p_\phi(\btheta\given\bfi{x})$ and $p(\btheta\given\bfi{x})$: 
$D_{\rm KL} (p\,||\,p_\phi)$.

In practice, we split the training data into a training and validation set with
a 90/10 split. 
Afterwards, we maximize the total log likelihood 
$\sum_i \log p_\phi(\btheta_i\given \bfi{x}_i)$ over training set, which is
equivalent to minimizing $D_{\rm KL} (p\,||\,p_\phi)$.
We use the {\sc Adam} optimizer~\citep{kingma2017} with a learning rate of $5\times10^{-4}$. 
To prevent overfitting, we evaluate the total log likelihood on the validation
data at every training epoch and stop the training when the validation log
likelihood fails to increase after 20 epochs.  
Training our model with a batch size of 50 takes roughly a day on a single 2.6
GHz Intel Skylake CPU. 
Given our small batch size, we find similar training times when using CPUs or
GPUs. 

\section{NASA-Sloan Atlas} \label{sec:obs}
As a demonstration of its speed and accuracy, we apply \sedflow~to optical
photometry from the NASA-Sloan Atlas\footnote{\url{http://nsatlas.org/}} (NSA).
The NSA catalog is a re-reduction of SDSS DR8~\citep{aihara2011} that includes
an improved background subtraction~\citep{blanton2011}.
We use SDSS photometry in the $u$, $g$, $r$, $i$, and $z$ bands, which are
corrected for galactic extinction using \cite{schlegel1998}.
To ensure that the galaxy sample is not contaminated, we impose a number of
additional quality cuts by excluding:
\begin{itemize}
    \item objects where the centroiding algorithm reports the
    position of the peak pixel in a given band as the centroid. 
    The SDSS photometric pipeline can struggle to accurately define the center
    of objects near the edge or at low signal-to-noise, so these cases are
    often spurious objects. 
    \item objects that have pixels that were not checked for peaks by the
    deblender. 
    \item objects where more than 20\% of point-spread function (PSF) flux is
    interpolated over as well as objects where the interpolation affected many
    pixels and the PSF flux error is inaccurate. 
    The SDSS pipeline interpolates over pixels classified as bad (\eg~cosmic
    ray).
    \item objects where the interpolated pixels fall within 3 pixels of their
    center and they contain a cosmic ray that was interpolated over.
    \item objects that were not detected at $\ge5\sigma$ in the original frame,
    that contain saturated pixels, or where their radial profile could not be
    extracted.
\end{itemize}
By excluding these objects, we avoid complications from artificats in the
photometry that we do not model. 
For additional details on the quality flags, we refer readers to the SDSS
documentation\footnote{\url{https://www.sdss.org/dr16/algorithms/flags_detail}}.
After the quality cuts, we have a total of 33,884 NSA galaxies in our sample.


\begin{figure}
\begin{center}
    \includegraphics[width=0.9\textwidth]{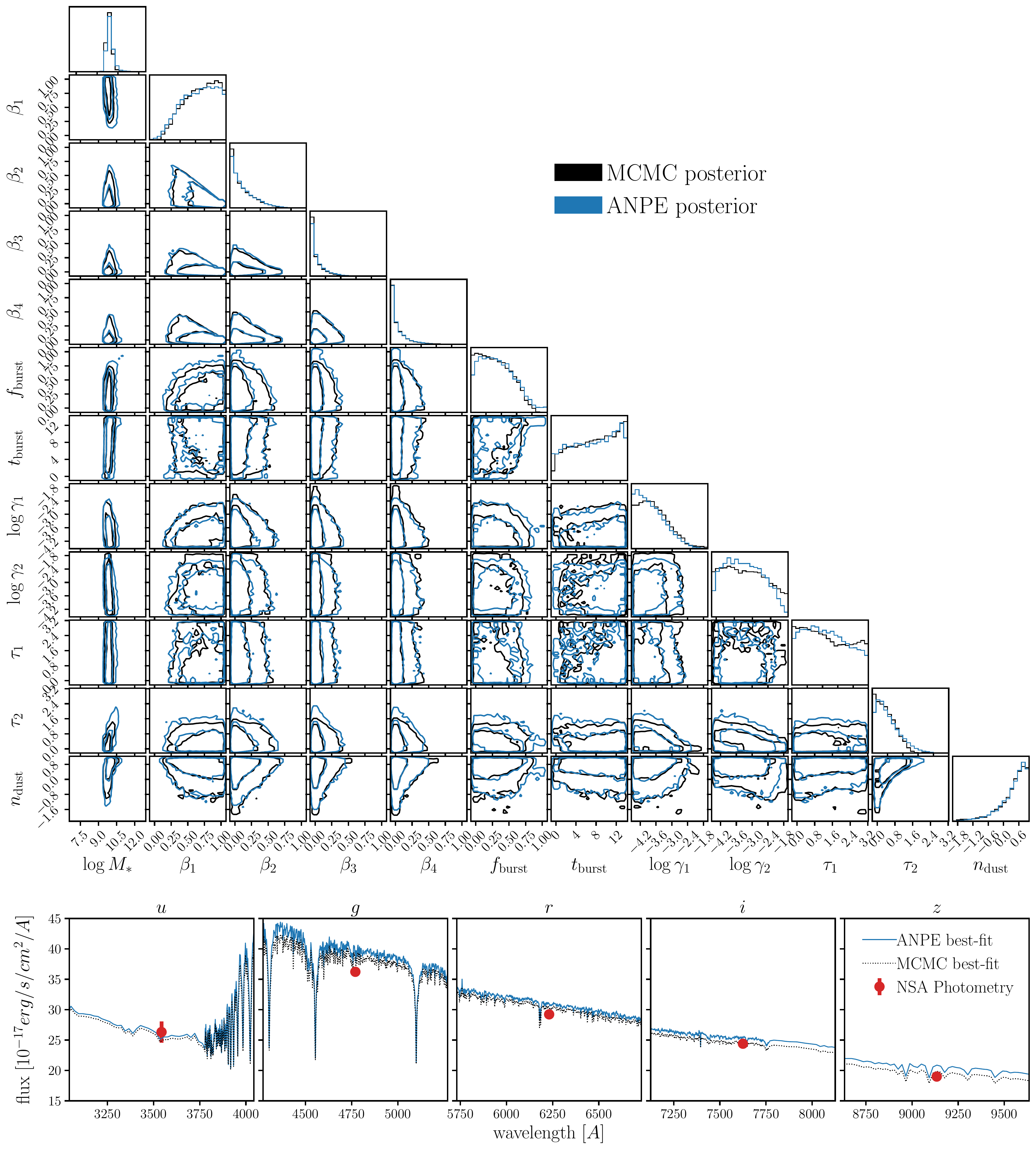}
    \caption{\label{fig:corner}
    A comparison of the posteriors of the 12 SED model parameters derived from
    standard MCMC sampling (black) and our \sedflow~(blue) for an arbitrarily
    selected NSA galaxy ($\mathtt{NSAID} = 72$; red in Figure~\ref{fig:data}).
    The posteriors are in excellent agreement for all of the SED parameters. 
    In the bottom panel, we present the SEDs of the the best-fit parameter
    values from the \sedflow~(blue) and MCMC posteriors (black dotted), which
    are both in good agreement with the observed NSA photometric flux (red). 
    Estimating the posterior using MCMC sampling requires $\sim$10 CPU hours. 
    Even using neural emulators to accelerate likelihood evaluations, MCMC
    sampling requires $\sim$10 CPU minutes. 
    \emph{With \sedflow, inferring the full posterior takes 1 second per
    galaxy.}
    }
\end{center}
\end{figure}

\section{Results} \label{sec:results}
Now that we have trained \sedflow, we can estimate the posterior,
$p(\btheta\given \bfi{x}_i)$, for any 
$\bfi{x}_i = \{f_{X,i}, \sigma_{X,i}, z_i \}$. 
In practice, we do this by drawing samples from the \sedflow~NDE model. 
Since we use a normalizing flow, this is trivial:
we generate samples from the target distribution of the normalizing flow,  a
multivariate Gaussian distribution in our case, then we transform the samples
using the bijective transformation in Eq.~\ref{eq:normflow} that we trained.
The transformed samples follow $p_\phi(\btheta\given \bfi{x}_i)$ and 
estimate the posterior, $p(\btheta\given \bfi{x}_i)$. 

Next, we validate the accuracy of the \sedflow~posterior estimates.
As a first test, we compare the posterior from \sedflow~to the posterior derived
from MCMC sampling for a single arbitrarily chosen NSA galaxy in
Figure~\ref{fig:corner} ($\mathtt{NSAID} = 72$). 
In the top, we present the the posterior distribution of the 12 SED model
parameters for the \sedflow~posterior (blue) and MCMC posterior (black). 
\emph{The \sedflow~posterior is in excellent agreement with the MCMC posterior
for all of the SED parameters}. 
 
In the bottom of Figure~\ref{fig:corner}, we compare the SEDs of the best-fit
parameter values from the \sedflow~(blue) and MCMC posteriors (black dotted). 
We also include the NSA photometric flux of the selected galaxy (red). 
The best-fit SED from the \sedflow~posterior is in good agreement with
both the MCMC best-fit SED and the NSA photometry.  

The key advantage of ANPE is that it enables accurate Bayesian inference
orders of magnitude faster than conventional methods. 
We derive the MCMC posterior using the {\sc Zeus} ensemble
slice-sampler~\citep{karamanis2020} with 30 walkers and 10,000 iterations.
2,000 of the iterations are discarded for burn-in. 
In total, the MCMC posterior requires >100,000 SED model evaluations. 
Since each evaluation takes ${\sim}340$ ms, it takes ${\sim}10$ CPU hours for a
single MCMC posterior. 
Recently, SED modeling has adopted neural emulators to accelerate SED model
evaluations~\citep{alsing2020}. 
In \cite{hahn2022}, for instance, the PROVABGS emulator takes
only ${\sim}2.9$ ms to evaluate, >100$\times$ faster than the original model. 
Yet, even with emulators, due to the number of evaluations necessary for
convergence, an MCMC posterior takes ${\sim}10$ CPU minutes. 
Meanwhile, after training, \emph{the \sedflow~posterior takes $1$ second ---
>$10^4\times$ faster than MCMC}. 

\begin{figure}
\begin{center}
    \includegraphics[width=0.45\textwidth]{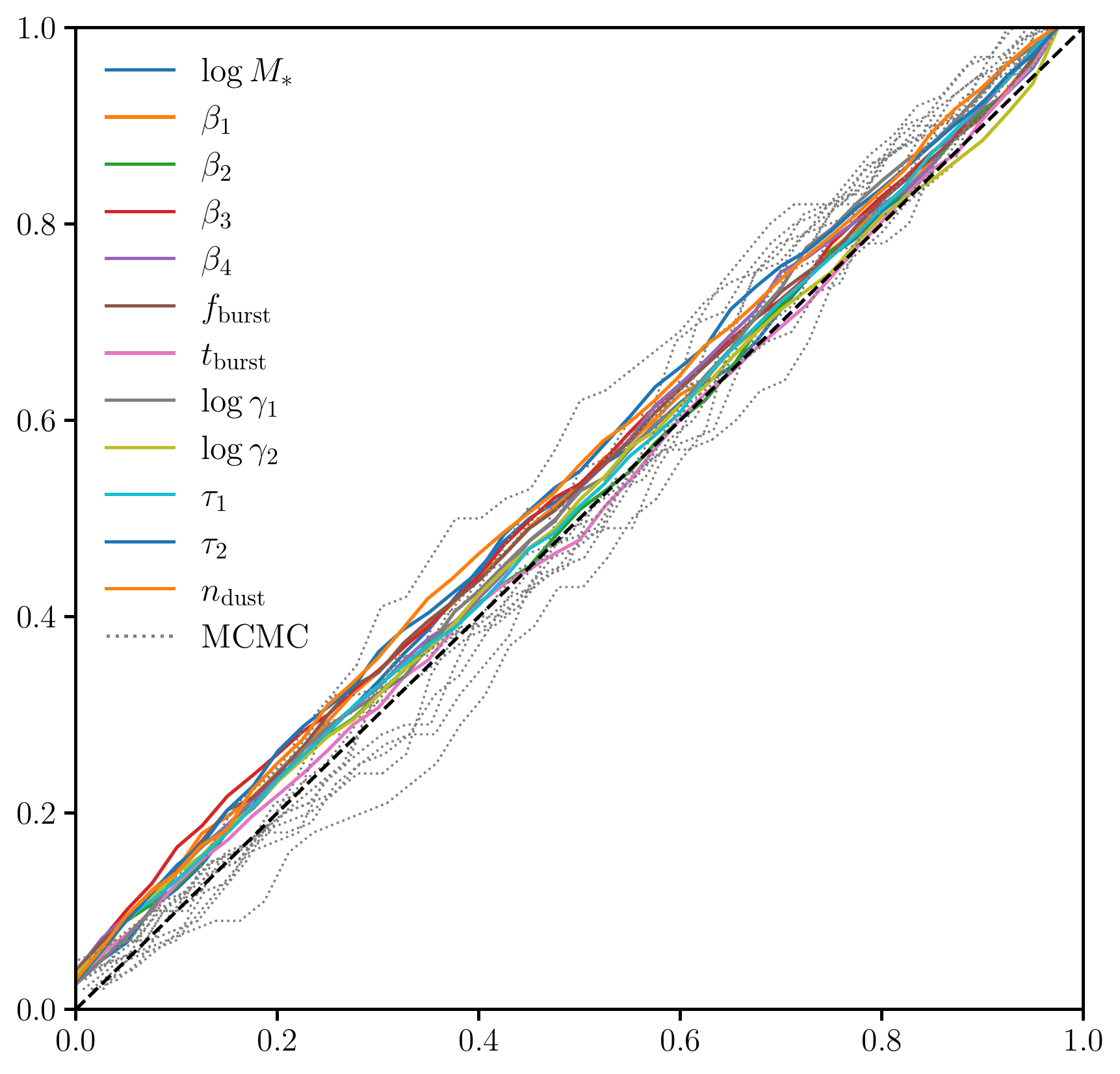}
    \caption{\label{fig:pp}
    Probability-probability (p-p) plot of the \sedflow~posteriors for 1000
    synthetic test observations, with known true parameter values. 
    We plot the CDFs of the percentile score of the true values within the
    marginalized \sedflow~posteriors for each SED parameter.
    For the true posteriors, the percentile score is uniformly distributed so
    the CDF is diagonal (black dashed).
    For reference, we include the p-p plot of the posterior estimated from MCMC
    sampling (gray). 
    \emph{The \sedflow~posteriors are in excellent agreement the true
    posteriors.}
    }
\end{center}
\end{figure}
The posteriors from \sedflow~and MCMC are overall in excellent for NSA
galaxies, besides the one in Figure~\ref{fig:corner}.
However, we do not know the true SED parameters for these galaxies so to
further validate \sedflow, we use test synthetic photometry, where we know the
truth.
We sample 1000 SED parameters from the prior,
$\{\btheta^{\rm test}_i\} \sim p(\btheta)$, 
and forward model synthetic NSA observations, 
$\{\bfi{x}^{\rm test}_i\}$, 
for them in the same way as the training data (Section~\ref{sec:training}). 
Afterwards, we generate posteriors for each of $\bfi{x}^{\rm test}_i$ using 
\sedflow: $\{ p(\btheta \given \bfi{x}^{\rm test}_i)\}$. 

In Figure~\ref{fig:pp}, we present the probability-probability (p-p) plot of
the \sedflow~posteriors for the test data. 
The p-p plot presents the cumulative distribution function (CDF) of the
percentile score of the true value within the marginalized posterior for each
parameter. 
For true posteriors, the percentiles are uniformly distributed so the CDF is a
diagonal (black dashed).
Overall, the CDFs for \sedflow~lie close to the diagonal for each of the SED
parameters. 
\emph{Hence, the \sedflow~posteriors are in excellent agreement with the true
posteriors}.

In Figure~\ref{fig:pp}, we also include the CDFs of the SED parameters for the
MCMC posteriors derived for a subset of 100 test observations (gray dotted). 
Comparing the CDFs from the MCMC posteriors to those of \sedflow, we find that
the \sedflow~posteriors are actually in better agreement with the true
posteriors. 
This is due to the fact that MCMC posteriors are also only estimates of the
true posterior and are subject to limitations in initialization, sampling, and
convergence.
Posteriors from \sedflow~are not impacted by these limations, so the comparison
highlights additional advantages of an ANPE approach besides the
${>}10^4\times$ performance improvement.

\begin{figure}
\begin{center}
    \includegraphics[width=0.85\textwidth]{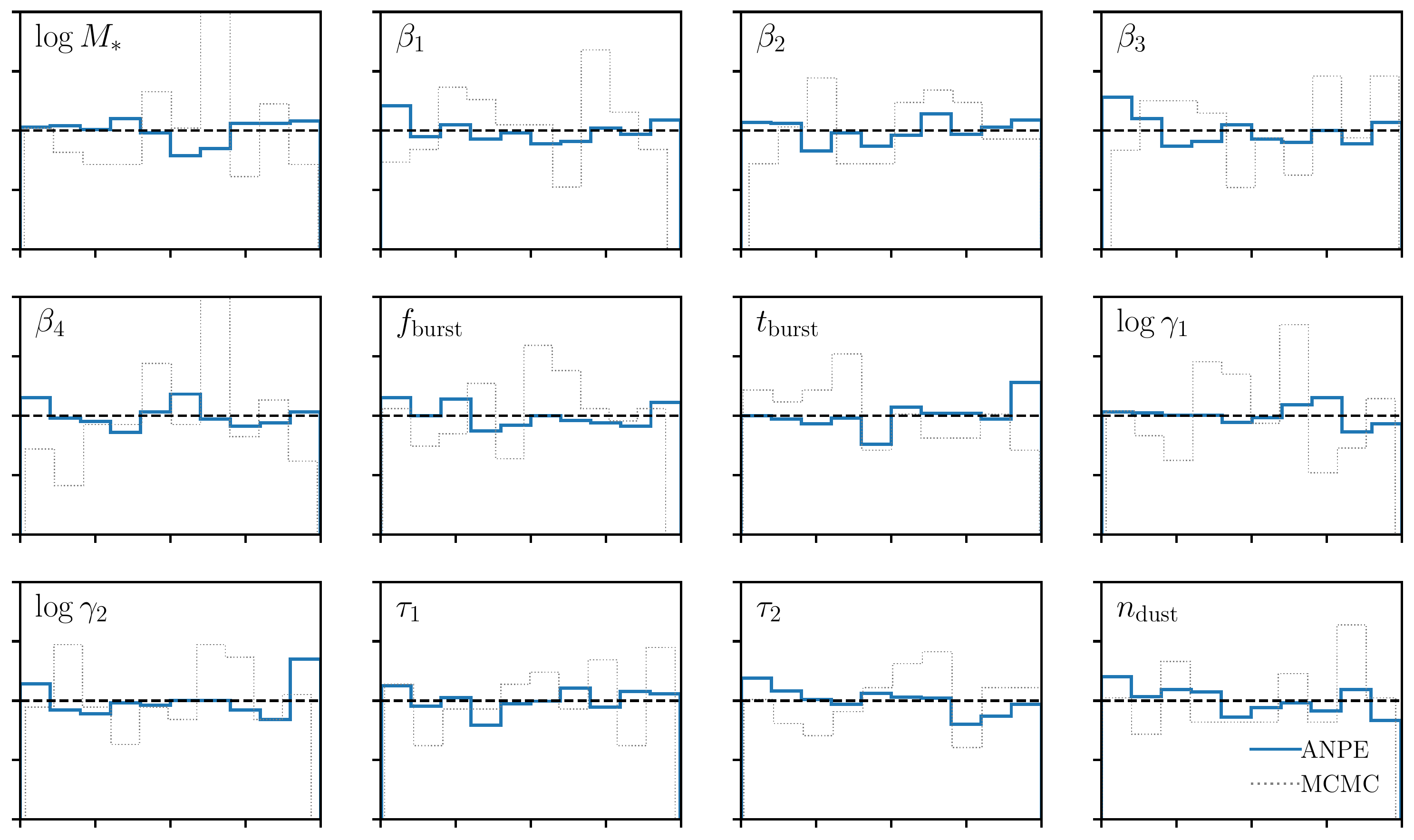}
    \caption{\label{fig:sbc}
    Simulation-based calibration plot of the \sedflow~posteriors for 1000
    synthetic test observations. 
    The histogram in each panel represents the distribution of the rank
    statistic of the true value within the marginalized \sedflow~posterior
    (blue) for each SED parameter.
    For the true posterior, the rank statistics will have a uniform
    distribution (black dashed). 
    For reference, we include the rank distribution of the MCMC posteriors for
    a subset of 100 test data (gray dotted). 
    The rank statistic distribution of \sedflow~is nearly uniform for all of
    the SED parameters. 
    Therefore, \emph{\sedflow~provides unbiased and accurate estimates of the
    true posteriors.}
    }
\end{center}
\end{figure}

We examine another validation of the \sedflow~posteriors using simulation-based
calibration~\citep[SBC;][]{talts2020}. 
Rather than using percentile scores, SBC examines the distribution of the rank
statistics of the true parameter values within the marginalized posteriors. 
It addresses the limitation that the CDFs only asymptotically approach the true
values and that the discrete sampling of the posterior can cause artifacts in
the CDFs. 
In Figure~\ref{fig:sbc}, we present SBC of each SED parameter for the
\sedflow~posteriors (blue) using the 1000 test observations.
For comparison, we include the SBC for the MCMC posteriors (gray dotted). 
Similar to the percentile score, the distribution of the rank statistic is
uniform for the true posterior (black dashed). 
The rank statistic distribution for the \sedflow~posteriors are nearly uniform
for all SED parameters. 
Hence, \emph{we confirm that the \sedflow~posteriors are in excellent agreement
with the true posterior}.

An advantage of SBC is that by examining the deviation of rank statistics
distribution from uniformity, we can determine how the posterior estimates
deviate from the true posteriors. 
For instance, if the distribution has a U-shape where the true parameter values
are more often at the lowest and highest ranks, then the posterior estimates
are narrower than the true posteriors.
If the distribution has a $\cap$-shape, then the posterior estimates are
broader than the true posteriors. 
Any asymmetry in the distribution implies that the posterior estimates are
biased.  
For the \sedflow~posteriors, we find none of these features for any of the SED
parameters. 
Hence, \sedflow~provides unbiased and accurate estimates of the true posteriors
for all SED parameters. 

With the accuracy \sedflow~validated, we apply it to derive posteriors for all
of our NSA galaxies (Section~\ref{sec:obs}). 
Analyzing all 33,884 NSA galaxies takes $\sim$12 CPU hours. 
For each galaxy, we generate 10,000 samples of the 12-dimensional posterior,
$p(\btheta \given \bfi{x})$. 
These posteriors on SED parameters represent posteriors on stellar mass, SFH,
ZH, and dust content of the NSA galaxies. 
To maximize the utility of the posteriors further, we use them to derive
posteriors on the following additional galaxy properties: SFR averaged over
1Gyr, mass-weighted metallicity, and mass-weighted stellar age (see
Eq. 17 in \citealt{hahn2022} for the exact calculation).
We publicly release all of the posteriors for our NSA sample as well as all of
the software and data used to train and validate \sedflow~at 
\url{https://changhoonhahn.github.io/SEDflow/}.

\section{Discussion} \label{sec:discuss}
\subsection{Forward Model} \label{sec:forward-model}
In the previous section, we demonstrated the accuracy of \sedflow~posteriors. 
Nevertheless, a primary determining factor for the fidelity of \sedflow, or any
ML model, is the quality of the training data set and, thus, the forward model
used to construct it. 
Below, we discuss the caveats and limitations of our forward model, which has
two components: the PROVABGS SPS model and noise model 
(Section~\ref{sec:training}).

First, for our noise model, we assign uncertainties to noiseless photometric
fluxes based on an empirical estimate of $p(\sigma_X\given f_X)$ for each band
independently. 
This is a simplicistic prescription and, as the bottom right panels of
Figure~\ref{fig:data} ($g - \sigma_r$ and $r - \sigma_r$) reveal, there are
discrepancies between the magnitude - uncertainty distributions of the training
data and observations. 
Despite these discrepancies, \sedflow~provides excellent estimates of the true
posterior.  
This is because we design our ANPE to include $\sigma_X$ as a conditional
variable (Section~\ref{sec:anpe_train}).
The $f_X-\sigma_X$ distribution of our training data does not impact the
accuracy of the posteriors as long as there are sufficient training data near
$\bfi{x}$ to train the NDE in that region.

A more accurate noise model will, in theory, improve the performance of
\sedflow~because the $\bfi{x}$-space of the training data will more efficiently 
span the observations. 
Fewer training data would be expended in regions of $\bfi{x}$-space that are
devoid of observations.  
However, for our application, we do not find significantly  
improved performance when we alter the noise model.
This suggests that even with our simplistic noise model, the $\bfi{x}$-space of
observations is covered sufficiently well by the training data. 
We note that when we decrease $N_{\rm train}$ to below 500,000,
\sedflow~posteriors are significantly less accurate. 
A more realistic forward model may reduce this $N_{\rm train}$ threshold for
accurate posteriors. 
However, generating $N_{\rm train}{\sim}1,000,000$ training SEDs has a
negligible computational cost compared to MCMC SED modeling, so we do not
consider it necessary to explore this further. 

Next, we consider limitations in the PROVABGS SPS model used in our forward
model. 
Our SPS model uses a compact and flexible prescription for SFH and ZH that can
describe a broad range of SFHs and ZHs.
However, the prescription is derived from simulated Illustris galaxies, whose
SFHs and ZHs may be not reflect the full range of SFHs and ZHs of real
galaxies.
If certain subpopulations of observed galaxies have SFHs and ZHs that cannot be
well described by the PROVABGS prescription, they cannot be accurately modeled.
Even if the PROVABGS SFH and ZH prescriptions are sufficient, there are
limitations in our understanding of stellar evolution. 

There is currently no consensus in the stellar evolution, stellar spectral
libraries, or IMF of galaxies~\citep[\emph{e.g.}][]{treu2010, vandokkum2010,
rosani2018, ge2019, sonnenfeld2019}.
The PROVABGS model uses MIST isochrones, \cite{chabrier2003} IMF, and the MILES
+ BaSeL spectral libraries. 
These choices limit the range of SEDs that can be produced by the training
data. 
For instance, if galaxies have significant variations in their IMF, assuming a
fixed IMF would falsely limit the range of our training data.  
A more flexible SED model that includes uncertainties in SPS would broaden the
range of galaxy SEDs that can be modeled.
Data-driven approaches may also enable SED models to be more
descriptive~\citep[\emph{e.g.}][]{hogg2016, portillo2020}. 
Improving  SED models, however, is beyond the scope of this work. 
Our focus is on improving the Bayesian inference framework.
In that regard, the limitations of the SED model equally impacts conventional
approaches with MCMC. 

We encounter the caveats above when we apply \sedflow~to the NSA catalog. 
For a small fraction of NSA galaxies (588 out of 33,884), \sedflow~generates
posteriors that are outside of the prior volume. 
This is because the photometry or uncertainties of these galaxies lie outside
of the support of the training data and where \sedflow~is well trained. 
They either have higher photometric uncertainties, for a given magnitude, or
bluer photometric colors than the training data. 
Some of these may be observational artifacts or problematic photometry.
Nevertheless, \sedflow~fails because we cannot construct training data near
them with our limited noise and SPS models. 
Since this only affects a small fraction of the NSA galaxies, we flag them in
our catalog and, for completeness, infer their galaxy properties by applying
PROVABGS with MCMC sampling.
For more details, we refer readers to Appendix~\ref{sec:fail}.

To test for limitations of the forward model, we can construct additional tests
of posteriors derived from ANPE. 
For instance, the $\chi^2$ of the best-fit parameter value from the 
estimated posterior can be used to assess whether the best-fit model 
accurately reproduces observations.
This would only require one additional model evaluation per galaxy. 
One can also construct an Amortized Neural Likelihood Estimator (ANLE) using
the same training data.
Unlike the ANPE, which estimates $p(\btheta\given f_X, \sigma_X, z)$, the ANLE
would estimate $p(f_X \given \btheta, \sigma_X, z)$.
We can then further validate the posteriors by assessing whether the observed
photometry lies within the ANLE distribution. 
Based on the overall high level of accuracy of \sedflow~posteriors, we do not
explore these additional tests; however, they can be used to further validate
any ANPE posteriors. 

\subsection{Advantages of \sedflow} 
The primary advantage of \sedflow~is its computational speed. 
This becomes even more pertinent if we want to add additional parameters to
address concerns about the choices in current SPS models, described above.
To relax these assumptions, SPS models would need to introduce additional
parameters that flexibly model these uncertainties~\citep{conroy2009,
conroy2010c}. 
While the dimensionality of current SPS models already makes MCMC methods
computational infeasible, ANPE has been applied to higher dimensional
applications.
For instance, \cite{dax2021} constructed an accurate ANPE for a
15-dimensional model parameter and 128-dimensional conditional variable
spaces.
NDE is an actively developing field in ML and new methods are constantly
emerging~\citep[\eg][]{wu2020, dhariwal2021}. 
Since ANPE can handle higher dimensionality, we can in the future include
additional parameters that model uncertainties in SPS. 
This will not only improve our SED modeling, but also improve our understanding
of stellar evolution and the IMF.

In addition to enabling scalable SED modeling for the next generation galaxy
surveys, \sedflow~will also enable us to tackle other key challenges in SED
modeling. 
For example, recent works have demonstrated that priors of SED models can
significantly impact the inferred galaxy properties~\citep{carnall2018,
leja2019, hahn2022}. 
Even ``uniformative'' uniform priors on SED model parameters can impose
undesirable priors on derived galaxy properties such as $M_*$, SFR, SFH, or
ZH.
To avoid significant biases, galaxy studies must carefully select priors and
validate their results using multiple different choices. 
With an MCMC approach, selecting a different prior means reevaluating every
posterior and repeating all the SED model evaluations in the MCMC sampling.  
For an ANPE approach, the prior is set by the distribution of parameters in the
training data. 
For a new prior, instead of reconstructing the training data, we can resample
it in such a way that the parameters follow the new prior.
Then, the ANPE model can be re-trained, re-validated on the test data, and
re-deployed on observations.
Each of these steps require substantially less computational resources than
generating a new set of training data or using MCMC methods. 
Hence, the ANPE approach provides a way to efficiently vary the prior without
multiplying computational costs.

\section{Summary and Outlook} \label{sec:summary}
By analyzing the SED of a galaxy, we can infer detailed physical properties
such as its stellar mass, star formation rate, metallicity, and dust content. 
These properties serve as the building blocks of our understanding of how
galaxies form and evolve. 
State-of-the-art SED modeling methods use MCMC sampling to perform Bayesian
statistical inference. 
They derive posterior probability distributions of galaxy properties given
observation that accurately estimate uncertainties and parameter degeneracies
to enable more rigorous statistical analyses. 
For the dimensionality of current SED models, deriving a posterior requires 
${\gtrsim}100,000$ model evaluations and take ${\gtrsim}10-100$ CPU hours per 
galaxy. 
Upcoming galaxy surveys, however, will observe \emph{billions} of galaxies
using \emph{e.g.} DESI, PFS, Rubin observatory, James Webb Space Telescope, and
the Roman Space Telescope. 
Analyzing all of these galaxies with current Bayesian SED models is infeasible
and would require hundreds of billions of CPU hours.
Even with recently proposed emulators, which accelerate model evaluations by
three to four orders to magnitude, the computation cost of SED modeling would
remain a major bottleneck for galaxy studies. 

We demonstrate in this work that Amortized Neural Posterior Estimation (ANPE)
provides an alternative \emph{scalable} approach for Bayesian inference in SED
modeling.
ANPE is a simulation-based inference method that formulates Bayesian inference
as a density estimation problem and uses neural density estimators (NDE) to
approximate the posterior over the full space of observations. 
The NDE is trained using parameter values drawn from the prior and mock
observations simulated with these parameters.  
Once trained, a posterior can be obtained from the NDE by providing the
observations as the conditional variables without any additional model
evaluations. 

In this work, we present {\sc SEDflow}, a galaxy SED modeling method using ANPE
and PROVABGS, a flexible SED model that uses a compact non-parameteric SFH and
ZH prescriptions and was recently validated in \cite{hahn2022}.
Furthermore, we apply {\sc SEDflow} to optical photometry from the NASA-Sloan
Atlas as demonstration and validation of our ANPE approach.  
We present the following key results from our analysis. \vspace{2mm}
\begin{compactitem}
    \item We train {\sc SEDflow} using a data set of ${\sim}1$ million SED
        model parameters and forward model synthetic SEDs.
        The parameters are drawn from a prior and the forward model is based on
        the PROVABGS and noise models. 
        We design the ANPE to estimate $p(\btheta | f_X, \sigma_X, z)$, where
        $f_X$, $\sigma_X$, and $z$ are the photometry, photometric uncertainty,
        and redshift, respectively. 
        For its architecture, we use a MAF normalizing flow with 15 MADE blocks
        each with 2 hidden layers and 500 hidden units.
        Training {\sc SEDflow} requires roughly 1 day on a single CPU. 
        Once trained, deriving posteriors of galaxy properties for a galaxy
        takes ${\sim}1$ second, $10^5\times$ faster than traditional MCMC sampling. 
    \item Posteriors derived using {\sc SEDflow} show excellent agreement with
        posteriors derived from MCMC sampling. 
        We further validate the accuracy of the posteriors by applying  {\sc
        SEDflow} to synthetic observations with known true parameter values.  
        Based on statistical metrics used in the literature (p-p plot and SBC),
        we find excellent agreement between the {\sc SEDflow} and the true
        posteriors. 
    \item Lastly, we demonstrate the advantages of {\sc SEDflow} by applying it
        to the NASA-Sloan Atlas.
        Estimating the posterior of ${\sim}33,000$ galaxies takes $\sim$12 CPU
        hours.
        We make the catalog of posteriors publicly available at
        \url{https://changhoonhahn.github.io/SEDflow/}. 
        For each galaxy, the catalog contains posteriors of all 12 PROVABGS
        SED model parameters as well as the galaxy properties: $M_*$, 
        average SFR over 1Gyr, mass-weighted metallicity, and mass-weighted
        stellar age. \vspace{2mm}
\end{compactitem}

This work highlights the advantages of using an ANPE approach to Bayesian SED
modeling. 
Our approach can easily be extended beyond this application. 
For instance, we can include multi-wavelength photometry at ultra-violet or
infrared wavelengths. 
We can also modify \sedflow~to infer redshift from photometry. 
In \sedflow, we include redshift as a conditional variable, since NSA provides
spectroscopic redshifts. 
However, redshift can be included as an inferred variable rather than a
conditional one. 
Then, we can apply \sedflow~to infer galaxy properties from photometric data
sets without redshift measurements while marginalizing over the redshift
prior. 
If we do not require spectroscopic redshifts, \sedflow~can be extended to much
larger data sets that span fainter and broader galaxy samples. 
Conversely, we can use \sedflow~to infer more physically motivated photometric 
redshifts, where we marginalize over our understanding of galaxies rather than
using templates. 

The ANPE approach to SED modeling can also be extended to galaxy spectra. 
Constructing an ANPE for the full data space of spectra would requires
estimating a dramatically higher dimensional probability distribution. 
SDSS spectra, for instance, have ${\sim}3,600$ spectral elements.  
In our approach we include the uncertainties of observables as conditional
variables, which doubles the curse of dimensionality.
Recent works, however, have demonstrated that galaxy spectra can be represented
in a compact low-dimensional space using autoencoders~\citep[][Melchior \&
Hahn, in prep.]{portillo2020}.
In \cite{portillo2020}, they demonstrate that SDSS galaxy spectra can be
compressed into 7-dimensional latent variable space with little loss of
information. 
Such spectral compression dramatically reduces the dimensionality of the
conditional variable space to dimensions that can be tackled by current ANPE
methods. 
We will explore SED modeling of galaxy spectrophotometry using ANPE and
spectral compression in a following work.

\section*{Acknowledgements}
It's a pleasure to thank 
    Adam Carnall, 
    Miles Cranmer, 
    Kartheik Iyer,
    Andy Goulding,
    Jenny E. Green,
    Jiaxuan Li, 
    Uro{\u s}~Seljak,
    and 
    Michael A. Strauss
for valuable discussions and comments.
This work was supported by the AI Accelerator program of the Schmidt Futures Foundation.

\appendix
\section{Testing Outside \sedflow~Training Range} \label{sec:fail}
\begin{figure}
\begin{center}
    \includegraphics[width=0.9\textwidth]{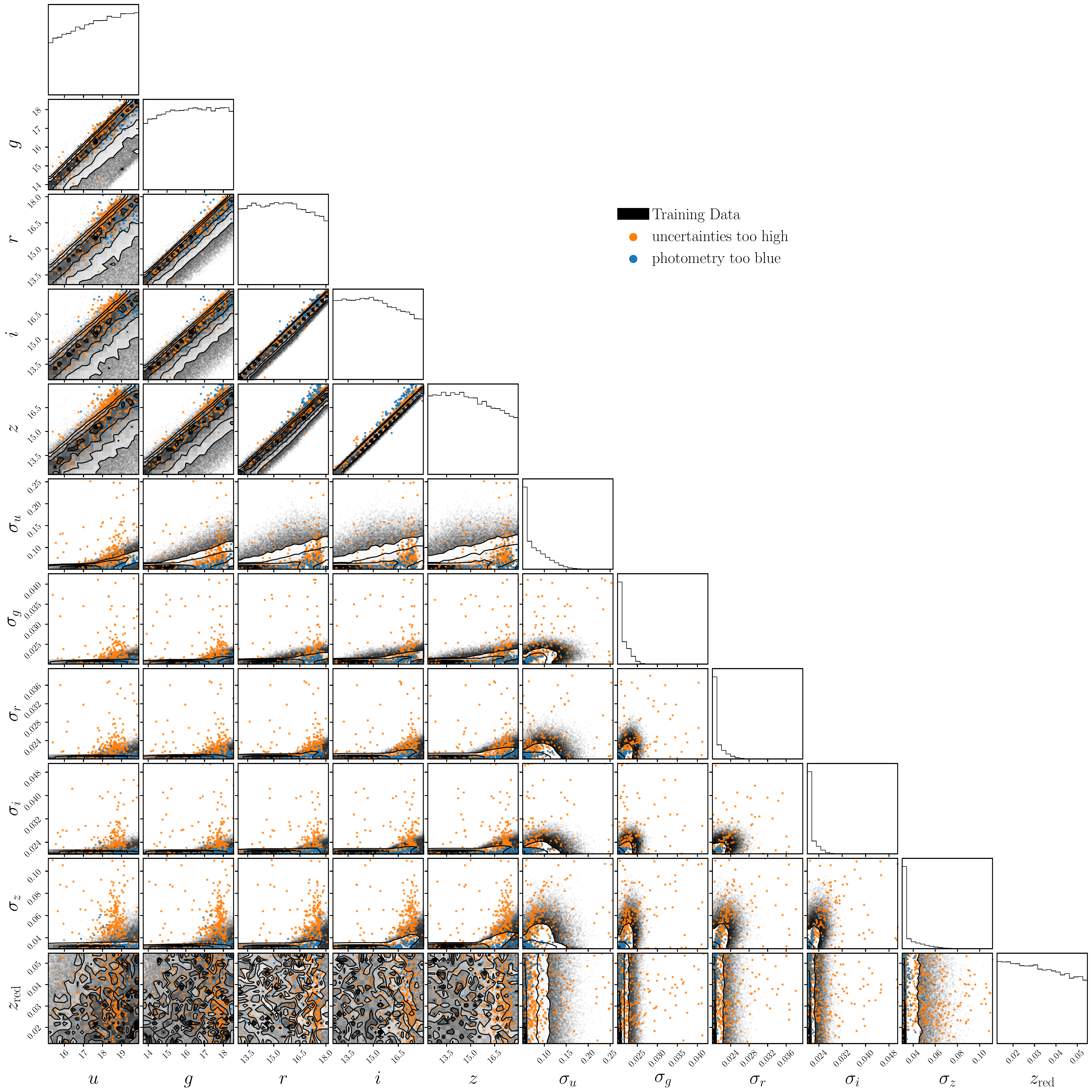}
    \caption{\label{fig:fail}
    The distribution of photometric magnitudes, uncertainities, and redshift of
    NSA galaxies for which \sedflow~does not produce valid posteriors (blue or
    orange). 
    These galaxies lie outside of the suport of the \sedflow~training data
    (black) so \sedflow~cannot accurately estimate their posteriors. 
    We mark the NSA galaxies that have unusually high photometric uncertainties
    that are not accounted for in our noise model in orange and galaxies that
    have photometric colors that cannot be modeled by our SED model in blue. 
    }
\end{center}
\end{figure}

For a small number of NSA galaxies, 588 out of 33,884, \sedflow~does not
produce valid posteriors. 
The normalizing flow of \sedflow~generates posteriors that are entirely outside
of the prior volume because the photometry or uncertainties of the galaxies lie
outside of the support of the \sedflow~training data
(Section~\ref{sec:training}). 
These galaxies either  
(1) have unusually high photometric uncertainties that are not accounted
for in our noise model or 
(2) they have photometric colors that cannot be modeled by our SED model. 
In Figure~\ref{fig:fail}, we present the distribution of photometric
magnitudes, uncertainties, and redshift ($\bfi{x} = \{f_X, \sigma_X, z\}$) of
these NSA galaxies. 
We mark galaxies that are outside the \sedflow~training data support (black)
due to (1) in orange and (2) in blue. 

We classify the galaxies as (1), if they have $\sigma_X$ that is unusually high
for a given $f_X$ in at least one photometric band: 
$\sigma_X > \mu_{\sigma_X}(f_X) + 3 \sigma_{\sigma_X}(f_X)$
(see Eq.~\ref{eq:noise}). 
There are 490 galaxies without valid \sedflow~posteriors due to (1).
Many of them lie well beyond the locus of training data points. 
The \sedflow~estimate of $p(\btheta \given f_X, \sigma_X, z)$ is only accurate
in regions of $\bfi{x}$-space where there is sufficient training data. 
This requirement is not met for these galaxies.  
In principle, if we use a more conservative noise model than Eq.~\ref{eq:noise}
and construct noisier training data, we can expand the support of \sedflow. 
\sedflow~would then produce sensible posteriors for more NSA galaxies. 
However, there is an inherent trade-off. 
For a training data set of fixed size, a more conservative noise model would
reduce the amount of training data in $\bfi{x}$-space where the vast majority
of NSA galaxies lie and can reduce the accuracy of the posteriors in these
regions. 
Since, \sedflow~fails for only a small fraction of NSA galaxies, we do not
explore more conserative noise models in this work.

Next, we examine the galaxies that lie outside of the \sedflow~support because
they have colors that cannot be modeled by our SED model. 
In Figure~\ref{fig:fail}, we classify NSA galaxies as (2) if any of their
colors (\emph{e.g.} $u-r$, $u-g$, $r-z$) is bluer than the 99.9\% percentile of
the training data color. 
There are 98 galaxies without valid \sedflow~posteriors due to (2).
The $i$ versus $z$ magnitude panel in particular highlights how a significant
number of galaxies are bluer than the training data. 
The fact that the training data do not span these colors suggests that the
PROVABGS SED model may not fully describe all types of galaxies in observations. 
As we discuss in Section~\ref{sec:discuss}, this may be due to limitations in
the SFH and ZH prescription in the PROVABGS model or our understanding of
stellar evolution. 
Limitations of the SED model equally impacts conventional MCMC sampling
approaches and is beyond the scope of this work. 
We, therefore, do not examine the issue further. 
For completeness, we derive posteriors for NSA galaxies, for which
\sedflow~fails, using PROVABGS with MCMC sampling with the same configuration
as \cite{hahn2022}. 

\bibliography{sedflow} 
\end{document}